\documentclass[twocolumn,superscriptaddress,showpacs,preprintnumbers,amsmath,amssymb,prl]{revtex4}

\usepackage{graphicx}
\usepackage{dcolumn}
\usepackage{bm}

\begin{document}


\title{Electrical measurement of a two-electron spin state in a double quantum dot}

\author{Nobuhiko Yokoshi}
\affiliation{JST-CREST, 4-1-8 Honcho, Saitama 332-0012, Japan}
\affiliation{Nanotechnology Research Institute, AIST, 1-1-1 Umezono, Tsukuba 305-8568, Japan}
\author{Hiroshi Imamura}
 \email{h-imamura@aist.go.jp}
\affiliation{Nanotechnology Research Institute, AIST, 1-1-1 Umezono, Tsukuba 305-8568, Japan}
\affiliation{JST-CREST, 4-1-8 Honcho, Saitama 332-0012, Japan}
\author{Hideo Kosaka}
\affiliation{Laboratory for Nanoelectronics and Spintronics, Research Institute of Electrical Communication, Tohoku University, Sendai 980-8577, Japan}
\affiliation{JST-CREST, 4-1-8 Honcho, Saitama 332-0012, Japan}

\date{\today}

\begin{abstract}
We propose a scheme for electrical measurement of two-electron spin states in a semiconductor double quantum dot. We calculated the adiabatic charge transfer when surface gates are modulated in time. Because of spin-orbit coupling in the semiconductor, spatial displacement of the electrons causes a total spin rotation. It follows that the expectation value of the transferred charge reflects the relative phase as well as the total spin population of a prepared singlet-triplet superposition state. The precise detection of the charge transfer serves to identify the quantum superposition. 
\end{abstract}

\pacs{73.21.La  03.67.-a  71.70.Ej}
\maketitle


Electron spins confined in semiconductor quantum dots provide intrinsic quantum bits (qubits)~\cite{Loss}. In 2002, an efficient framework for universal quantum computation using singlet-triplet qubits was proposed by Levy~\cite{Levy}. Two-electron states in a double quantum dot (DQD) are characterized by the charge number in each dot labeled $(m,n)$ and the total spin $s=\{S,T\}$. Then the basis set is composed of three singlet states $|S_{\pm} \rangle \equiv [|(2,0)S \rangle \pm |(0,2)S \rangle ]/\sqrt{2}$, $|S\rangle \equiv |(1,1)S\rangle$ and triplet states $|T_{\sigma} \rangle \equiv |(1,1)T_{\sigma}\rangle$ with a magnetic quantum number $\sigma=\{ 0,\pm 1 \}$. A single spin qubit is defined in one of the singlet-triplet subspaces, e.g., $S$-$T_0$ subspace. The proposal has inspired different experimental~\cite{Koppens, Petta, Nowack, Tarucha} and theoretical studies~\cite{TaylorNP,Hanson} in recent years.  

\begin{figure}[b]
\includegraphics[width=85mm]{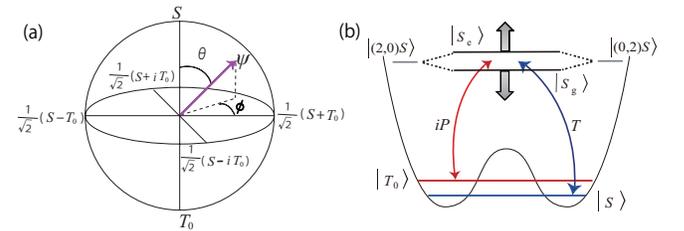}
\caption{(color online) (a) Bloch sphere of the $S$-$T_0$ spin subspace. The angle $\theta$ characterizes the total spin populations, whereas the angle $\phi$ denotes the relative phase. (b) Scattering processes in the DQD; $|S\rangle$ and $|T_0\rangle$ couple with each other through the doubly occupied states. $|S_{\rm g (e)}\rangle$ is a linear combination of $|(2,0)S\rangle$ and $|(0,2)S\rangle$, which diagonalizes the doubly occupied states when we set $T=P=0$ virtually. }
\label{model} 
\end{figure}

In the $S$-$T_0$ subspace, any of quantum superposition states can be written as
\begin{eqnarray}
|\psi\rangle
&=&\cos\frac{\theta}{2}|S\rangle+\sin\frac{\theta}{2} e^{i\phi}|T_0\rangle
\label{initial}
\end{eqnarray}
and is mapped on the Bloch sphere in Fig.~\ref{model}(a). To build the quantum gates, it is necessary to manipulate and detect the Bloch angles $\theta$ and $\phi$. However the conventional readout experiments using the Pauli spin blockade cannot detect the relative phase $\phi$, but can detect the total spin populations characterized by the angle $\theta$~\cite{Koppens, Petta, Nowack, Tarucha}. The relative phase $\phi$ is a fundamental element of the quantum mechanics in itself, and is essential in quantum algorithms such as Grover's database search problem~\cite{Grover}. Therefore we need to explore schemes for measuring the two Bloch angles in parallel.

In this work, we propose a measurement scheme for the coherence phase between the entangled spin states, which utilizes adiabatic charge transfer. It is assumed that the DQD defined by metal surface gates is located on a two-dimensional electron gas (2DEG) in a semiconductor. Then, we can control the electrical potentials in the two dots and the barrier potential separating them~\cite{Koppens, Petta,Waugh}. We calculated the charge difference in the DQD when the quantum state in Eq.~(\ref{initial}) is temporally varied by the gate manipulation. It was shown that the expectation value of the transferred charge depends on the initial phase $\phi$ due to spin-orbit couplings in the 2DEG. Thus, it becomes possible to identify the state vector on the Bloch sphere in the $S$-$T_0$ subspace by charge-sensing measurements at each dot~\cite{Koppens, Petta, Nowack, Tarucha, Field}. 

The system we consider is well described by the Hund-Mulliken model, in which doubly occupied states are taken into account. In this model, the orthonormalized single-particle state is defined as $\Phi_{\rm L(R)}=(\varphi_{\rm L(R)}-g\varphi_{\rm R(L)})/\sqrt{1-2\Gamma g+g^2}$ where $\Gamma$ is the overlap integral of the orbitals in the left and right dots $\varphi_{\rm L,R}$, and $g=(1-\sqrt{1-\Gamma^2})/\Gamma$~\cite{Burkard}. We consider the spin-orbit interactions for the singlet-triplet coupling~\cite{Stepanenko,Flindt,Ganga}. Electron tunneling thus accompanies the spin precession with respect to the vector ${\bm P}/2=-i\langle \Phi_{\rm L}|{\bm \Omega}|\Phi_{\rm R}\rangle$, in which the spin-orbit interaction is expressed by $\hat{H}_{\rm so}=\Sigma_{i=1,2} {\bm \Omega}({\bf k}_i)\cdot {\bf S}_i$ with ${\bf k}_i$ and ${\bf S}_i$ being the wave vector and spin operator of $i$-th excess electron, respectively~\cite{Stepanenko}. It is convenient to take the spin quantization axis to be parallel to ${\bm P}/2$. For this choice, the states $|T_{\pm 1}\rangle$ are decoupled from the other states~\cite{Stepanenko}, and will be omitted below. Thus, the Hamiltonian in $\{ |S_+\rangle, |S_-\rangle, |S\rangle, |T_0\rangle \}$ basis can be written as~\cite{Burkard,Stepanenko,Schliemann}
\begin{eqnarray}
\hat{H}=
\left(
\begin{array}{cccc}
U+X & \varepsilon & T & -iP \\
\varepsilon & U-X & 0 & 0 \\
T & 0 & 0 & 0 \\
iP & 0 & 0 & 0
\end{array}
\right),
\label{hamiltonian}
\end{eqnarray}
in which $U=\langle\Psi_{\rm L(R)}|C|\Psi_{\rm L(R)}\rangle$ and $X=\langle\Psi_{\rm L}|C|\Psi_{\rm R}\rangle$ with $\Psi_{\rm L(R)}=\Phi_{\rm L(R)}({\bf r}_1)\Phi_{\rm L(R)}({\bf r}_2)$ and $C({\bf r}_1,{\bf r}_2)$ being a Coulomb interaction. $T/2$ and $\varepsilon$ are the single-particle interdot tunneling and the gate-controlled potential difference between the dots, respectively. One can see that the singlet and the triplet $(1,1)$ states couple with each other through the doubly occupied states [see Fig.~\ref{model}(b)]. We assume that $\Gamma$ and $\varepsilon$ are the accessible parameters by modulating the gate voltages. The matrix elements $T$ and $P$ are approximately proportional to $\Gamma$~\cite{Burkard,Stepanenko}, and $X$ to $\Gamma^2$~\cite{Burkard} in the tunneling regime.

In Eq.~(\ref{hamiltonian}), only the ground orbital of each dot is considered. When one of the electrons is brought into the first excited orbital, the additional singlet and triplet states $\{|(0,2)S'\rangle, |(2,0)S'\rangle, |(0,2)T_{\sigma}\rangle, |(2,0)T_{\sigma}\rangle \}$ are possible~\cite{Hanson}. However, in the typical experiments, these states lie far ($\gtrsim 0.4 {\rm meV} \gg T,P$) above the $|S_{\pm}\rangle$ state~\cite{Koppens, Petta, Nowack, Tarucha}. Thus we can disregard them as long as $\varepsilon \lesssim U$.

We performed a unitary transformation of $|S_{\pm}\rangle$ so that, as $|\varepsilon|$ increases, one of them $|S_{\rm g}\rangle$ energetically approaches the $(1,1)$ states while the other state $|S_{\rm e}\rangle$ draws apart [see Fig.~\ref{model}(b)]. After adiabatic elimination of the higher state $|S_{\rm e}\rangle$~\cite{ae}, the effective Hamiltonian is given by
\begin{eqnarray}
\hat{H}_{\rm eff}
=
\left(
\begin{array}{ccc}
E_{\rm g} & T \sin\eta & -iP \sin\eta \\
T \sin\eta & -\frac{T^2}{E_{\rm e}}\cos^2\eta & i\frac{T P}{E_{\rm e}}\cos^2\eta \\
iP  \sin\eta & -i\frac{T P}{E_{\rm e}}\cos^2\eta & -\frac{P^2}{E_{\rm e}}\cos^2\eta
\end{array}
\right),
\label{effectiveH}
\end{eqnarray}
where $E_{\rm g(e)}=U \mp \sqrt{X^2+\varepsilon^2}$ and $\tan \eta=(-X+\sqrt{X^2+\varepsilon^2})/\varepsilon$. Here we introduce the instantaneous eigenstates and energies of the time-dependent Hamiltonian of Eq.~(\ref{effectiveH}), such that $\hat{H}_{\rm eff}(t)|m(t)\rangle=E_m(t)|m(t)\rangle$. Due to the presence of the spin-orbit couplings, the $(1,1)$ charge state is expanded with the so-called bright state $|{\rm B}\rangle = \cos\Theta |S \rangle + i \sin \Theta |T_0\rangle$ and dark state $|{\rm D}\rangle =\sin\Theta|S\rangle -i\cos\Theta|T_0\rangle$ with $\tan\Theta=P/T$~\cite{Bergmann}. Therefore one sees that the instantaneous eigenstates are $|{\rm D}\rangle$ and
\begin{eqnarray}
\begin{aligned}
|+\rangle &=~~\cos\Phi|S_{\rm g}\rangle+\sin\Phi|{\rm B}\rangle, \\
|-\rangle &=-\sin\Phi|S_{\rm g}\rangle+\cos\Phi|{\rm B}\rangle,
\end{aligned}
\label{eigen}
\end{eqnarray}
in which 
\begin{eqnarray}
&&\tan\Phi=\frac{\tilde{E}_{\rm g} +\sqrt{\tilde{E}_{\rm g}^2+4(T^2+P^2)\sin^2\eta}}{\sqrt{4(T^2+P^2)}\sin\eta}, \\
&&\tilde{E}_{\rm g}=E_{\rm g}+\frac{(T^2+P^2)}{E_{\rm e}}\cos^2\eta.
\end{eqnarray}
The dark state $|{\rm D}\rangle$ is free from the double occupancy state, and is not affected by the sweep of the bias potential. In addition, since the spin-orbit coupling is always weak compared with the hopping energy, the variation of the mixing angle $\Theta$ by the center gate is considerably small. On the other hand the angle $\Phi$ changes from $\pi/2$ to $0$ with increasing $\varepsilon$, which indicates that there is an avoided crossing of the $|{\rm B}\rangle$ and $|S_g\rangle$ states. The corresponding eigenenergies are respectively
\begin{eqnarray}
\begin{aligned}
E_{\rm D}&=0, \\
E_{\pm}&=\frac{\tilde{E}_{\rm g} \mp \sqrt{\tilde{E}_{\rm g}^2+4(T^2+P^2)\sin^2\eta}}{2} 
\\
&~~~~~~~~~~~~~~~~~~~~~~~~~~~~~~~
-\frac{(T^2+P^2)}{E_{\rm e}}\cos^2\eta.
\end{aligned}
\end{eqnarray}

We calculate the charge difference in the DQD after varying the system parameters adiabatically. The expected value of the charge difference is defined by 
\begin{eqnarray}
Q(\tau)=\int_0^{\tau} dt \langle \psi(t)|\hat{I}'|\psi(t)\rangle,
\label{pump}
\end{eqnarray}
in which $\hat{I}'$ is $3\times 3$ the ``current'' operator. The current operator is obtained by mapping $\hat{I}=(i/\hbar) [\hat{H},\hat{\rho}_z]$ on $\{ |{\rm D}\rangle, |+\rangle, |-\rangle \}$ basis, where $\hat{\rho}_z=2e[|(2,0)S\rangle\langle (2,0)S|-|(0,2)S\rangle\langle (0,2)S|]$ and $[~,~]$ denotes commutation. Because the dark state $|{\rm D}\rangle$ is decoupled from the double occupancy state, it does not contribute to the current. Then the detection of $Q(\tau)$ corresponds to the projection measurement in the bright state $|{\rm B}\rangle$. Note that in the Pauli spin blockade measurement the projection axis is $|S\rangle$~\cite{Koppens, Petta, Nowack, Tarucha}. Since the bright state lies on the sphere with a well-defined azimuthal angle of $\phi_{\rm B}=\pi/2$, the charge difference can capture the initial relative phase. 

With use of the instantaneous eigenstates, one can expand a two-electron state as 
\begin{eqnarray}
|\psi(t)\rangle=\sum_{m=\{ {\rm D},\pm \}} c_m(t) e^{i \zeta_m}|m;{\bf q}\rangle,
\label{expand}
\end{eqnarray}
where $\zeta_m(t)=-\int_0^t dt' E_m (t')/\hbar$ is the usual dynamical phase. The coefficient $c_m$ is varied with respect to the set of gate-controlled parameters ${\bf q}(t)=\{ \Gamma (t),\varepsilon (t)\}$, and obeys the differential equation~\cite{Pekola1,Pekola2}  
\begin{eqnarray}
\frac{dc_m}{dt}=-\sum_{n\not= m} c_n~e^{i\zeta_n- i\zeta_m}\langle m;{\bf q}|\frac{d}{dt}|n;{\bf q}\rangle.
\label{differential}
\end{eqnarray}
The time variation of $\delta c_m(t)=c_m(0)-c_m(t)$ represents the nonadiabatic level transition. However, as far as the adiabatic condition $|\langle m;{\bf q}|\hbar(d/dt)|n;{\bf q} \rangle/(E_m-E_n)| \ll 1$ is satisfied, it is negligible and may be dropped.
 
The adiabatically pumped charge difference is obtained by substituting Eqs.~(\ref{expand},\ref{differential}) into Eq.~(\ref{pump}), and taking the zero-order terms in terms of $\delta c_m(t)$. Here we assume that the manipulation time $\tau$ is much longer than the period of the unitary time evolution $e^{i\zeta_m}$. As a result, we find that the charge difference consists of two parts; $Q(\tau)=Q^0+Q^1$. One is 
\begin{eqnarray}
\hspace{-2mm}
Q^0
=
2\hbar{\rm Im} 
\sum_{m, n\not=m} |c_m(0)|^2 
\int_0^{\tau} dt 
\langle n|\frac{d}{dt}|m \rangle 
\frac{\langle m |\hat{I}'|n  \rangle}{E_m-E_n}.
\end{eqnarray}
The other is the interference part, which includes rapid inter-level oscillation in the integrand;
\begin{eqnarray}
Q^1
=2\hbar{\rm Im} \sum_{m,n\not=m}c_m(0)c_n^*(0) \int_0^{\tau} dt e^{i\zeta_m-i\zeta_n} 
K_{mn},
\end{eqnarray}
where
\begin{eqnarray}
K_{mn}(t)=
\frac{i}{2\hbar}\langle n|\hat{I}'|m \rangle 
- \sum_{l\not=\{ m,n \}} \langle m|\frac{d}{dt}|n \rangle 
\frac{\langle l|\hat{I}'|m \rangle}{E_l-E_m}.
\end{eqnarray}
When an eigenstate is prepared as an initial state, i.e., $c_m(0)=1$ for a certain $m$ and $c_{m'\not= m}(0)=0$, $Q(\tau)$ reduces to the result in the previous work~\cite{Pekola1,Pekola2}.

\begin{figure}[t]
\centering
\includegraphics[scale=0.28]{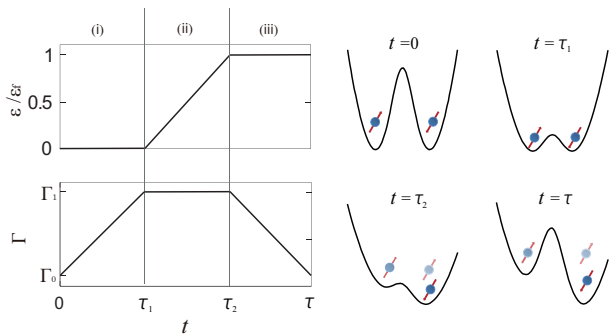}
\caption{
(color online) Sequence of gate voltage manipulations under consideration. The upper panel shows the potential bias between the left and the right dots. The overlap of the orbitals in the two dots $\Gamma$ is moved up and down by the gate-controlled potential barrier.
}
\label{sequence}
\end{figure}
Hereafter we calculate the charge difference $Q(\tau)$ for specific manipulation sequence. The gate voltages are initially adjusted so that no bias potential is present, and the barrier potential is so high that the system stays almost in the prepared state. The gate control under consideration is presented in Fig.~\ref{sequence}. The sequence consists of three parts; we (i) lower the interdot potential barrier in order to increase the overlap integral $\Gamma$, (i\hspace{-0.2mm}i) tilt the electric potential until $\varepsilon=\varepsilon_{\rm f}>0$ to be $\tilde{E}_{\rm g}(\varepsilon_{\rm f},\Gamma_0)=0$, and (i\hspace{-0.2mm}i\hspace{-0.2mm}i) raise the barrier potential height again.

Before proceeding to the calculation, we review the adiabatic conditions. First, no charge transfer occurs during the process (i) because the doubly occupied state is entirely decoupled. In the second process (i\hspace{-0.2mm}i), the gate sweep affects only the mixing angle $\Phi$. The dark state $|{\rm D}\rangle$ is thus left unperturbed. Then, the nonadiabatic contribution $\delta c_m$ is roughly proportional to the Landau-Zener (LZ) transition rate~\cite{LZ} between the doubly occupied state $|S_{\rm g}\rangle$ and the bright state $|{\rm B}\rangle$. At the avoided crossing, the LZ transition rate is estimated as
\begin{eqnarray}
p^{\rm (i\hspace{-0.2mm}i)}
\simeq \exp \Bigl\{ -\frac{\pi ( T_1^2+P_1^2 )}{\hbar |\dot{\varepsilon}|} \Bigr\},
\label{LZ1}
\end{eqnarray}
where $\dot{\varepsilon}=d\varepsilon /dt$, $T_{0(1)}=T(\Gamma_{0(1)})$ and $P_{0(1)}=P(\Gamma_{0(1)})$. In the last process (i\hspace{-0.2mm}i\hspace{-0.2mm}i), lowering $\Gamma$ changes both the mixing angles $\Theta$ and $\Phi$, but generally $d\Theta/dt \ll d\Phi/dt$.  Thus, assuming that the applied bias is so large that $\eta_{\rm f}(t)\equiv \eta(\varepsilon_{\rm f},\Gamma)\sim\pi/4$, the LZ transition rate is
\begin{eqnarray}
p^{\rm (i\hspace{-0.2mm}i\hspace{-0.2mm}i)}
\simeq \exp \Bigl\{ -\frac{\pi E_{\rm e}(\varepsilon_{\rm f})^2}{\hbar |\dot{\Lambda}|} \Bigr\},
\label{LZ2}
\end{eqnarray}
where $\dot{\Lambda}=d\sqrt{T^2+P^2} /dt$ is approximately proportional to $d\Gamma/dt$. It should be noted that the spin-orbit interaction appears in parallel with the interdot tunneling as long as the dark state is robust against the gate control. The coherent oscillation between the singlet and the triplet states holds when the adiabatic condition for the interdot tunneling $T/2$ is satisfied. This is a definite difference between the previous work using a field gradient for singlet-triplet mixing~\cite{Koppens, Petta, Nowack, Tarucha, TaylorNP} and ours. Besides, the single level spacing for the individual electrons is of the order of $1{\rm meV} \gg \sqrt{T^2+P^2}$. Therefore, the nonadiabatic transition to the excited orbitals is negligible as long as $p^{\rm (i\hspace{-0.2mm}i)}, p^{\rm (i\hspace{-0.2mm}i\hspace{-0.2mm}i)}\ll 1$ is satisfied. 

The target state we focus on is arbitrary superposition state of $|S\rangle$ and $|T_0\rangle$ as shown in Eq.~(\ref{initial}). At the initial condition $t=0$, it does not contain the excited state $|-\rangle$, i.e., $c_-(0)=0$. Therefore one can see that the interference part becomes
\begin{eqnarray}
Q^1 \propto \int_0^\tau dt  \frac{e^{i\zeta_+}\sin\Phi}{\sqrt{\tilde{E}_{\rm g}^2+4(T^2+P^2)\sin^2\eta}} \frac{d}{dt}\Theta.
\end{eqnarray}
As is mentioned above, the integrand is negligibly small and rapidly oscillating. Thus all we have to calculate is $Q^0$, which is proportional to $|c_+(0)|^2$. Within the first order of $P/T$, we obtain the adiabatically pumped charge difference as
\begin{eqnarray}
Q(\tau)
=-\frac{e}{2} 
\Bigl\{
(1+\cos\theta)+2\frac{P_0}{T_0}\sin\theta\sin\phi
\Bigr\}.
\label{pcharge}
\end{eqnarray}
The first term in the right-hand side describes the Pauli spin blockade~\cite{Petta,Nowack,Ono,Koppens}. It should be noted that the charge difference oscillates with respect to the relative phase $\phi$. The oscillating term reflects the imaginary part of $|\psi\rangle$, and originates from the projection axis $|{\rm B}\rangle$ which is slightly tilted from $|S\rangle$. In addition, the oscillation does not appear when the initial state is a pure singlet ($\theta=0$) or triplet state ($\theta=\pi$). 

For a clean GaAs/AlGaAs 2DEG confined in $10\sim 100 {\rm nm}$ long, the spin-orbit interaction energy is estimated to be $\langle \hat{H}_{\rm so} \rangle= 10^{-2}\sim 10^{-1} {\rm meV}$ from magnetoresistance data~\cite{Miller}. On the other hand the confinement energy is $\sim 1 {\rm meV}$ for a quantum dot with a $30{\rm nm}$ side~\cite{Burkard}. In that case, it is possible to experimentally achieve the condition in which the oscillation amplitude is $\sim 10\%$ of $Q(\tau,\phi=0)$~\cite{Stepanenko}. Therefore the repetitive experiments can reveal the relative phase as well as the total spin population of the prepared state.

So far, we have neglected the effect of nuclear spins in the semiconductor. The hyperfine fields due to the nuclei (Overhauser field) ${\bf h}_{\rm L(R)}$ couple with the electron spins as $V_{\rm hf}={\bf h}\cdot({\bf S}_{1}+{\bf S}_{2})+\delta{\bf h}\cdot({\bf S}_{1}-{\bf S}_{2})$, where ${\bf h}=({\bf h}_{\rm L}+{\bf h}_{\rm R})/2$ and $\delta{\bf h}=({\bf h}_{\rm L}-{\bf h}_{\rm R})/2$. Then the average of the Overhauser fields ${\bf h}$ rotates the subspace of the three spin triplet states, while the inhomogeneity $\delta{\bf h}$ mixes $|S\rangle$ with $|T_{\sigma}\rangle$s~\cite{Taylor,Coish,Cakir}. 

The Overhauser fields can disturb the electron spin state during the adiabatic gate control. However the hyperfine coupling between $|S\rangle$ and $|T_{\sigma}\rangle$ does not undergo virtual double occupancy, and the adiabatic condition for $\delta {\bf h}$ is different from that for $T/2$ and $P/2$. Thus, we can separate off the effect of the Overhauser field using the technique called ``rapid adiabatic passage'', in which the sweep of the bias is adiabatic for the electron tunneling but is nonadiabatic for the hyperfine couplings~\cite{Petta,Petta2}. In a quantum dot containing unpolarized $N=10^5$ nuclear spins, the root mean square of the Overhauser field is $|\langle {\bf h}_{\rm L(R)} \rangle_{\rm rms}| \sim 10^{-4} {\rm meV}$. The required length of the manipulation sequence $\tau$ is a few $\mu{\rm s}$ or shorter for interdot tunneling coupling $T_1/2\sim 10^{-2} {\rm meV}$. This condition has been achieved in a couple of experiments~\cite{Petta,Petta2}.

In summary, we propose an adiabatic charge transfer in a gate-defined DQD as an indicator of singlet-triplet quantum superposition on a $S$-$T_0$ Bloch sphere. After the gate manipulations, the transferred charge number is found to oscillate with respect to two Bloch angles in the initially prepared superposition state. The oscillation can be observed in an ensemble average of charge-sensing measurements in each dot with quantum point contacts~\cite{Koppens, Petta, Field}. Recently Kosaka {\it et al}. demonstrated the quantum coherence transfer from light polarization to electron spin polarization in a quantum well~\cite{Kosaka,Kosaka2}. By applying this method, it becomes possible to prepare arbitrary $S$-$T_0$ superposition states in DQD. The present scheme can help to check whether a system is indeed prepared in the desired state. 

The authors would like to thank Y. Rikitake, K. Matsushita, J. Sato, T. Taniguchi and H. Ohtori for useful comments and discussions.


\begin{thebibliography}{0}
\bibitem{Loss} D. Loss and D. P. DiVincenzo, Phys. Rev. A {\bf 57}, 120 (1998).
\bibitem{Levy} J. Levy, Phys. Rev. Lett. {\bf 89}, 147902 (2002).
\bibitem{Petta} J. Petta, A. C. Johnson, J. M. Taylor, E. A. Laird, A. Yacoby, M. D. Lukin, C. M. Marcus, M. P. Hanson, and A. C. Gossard, Science {\bf 309}, 2180 (2005).
\bibitem{Koppens} F. H. L. Koppens, J. A. Folk, J. M. Elzerman, R. Hanson, L. H. Willems van Beveren, I. T. Vink, H. P. Tranitz, W. Wegscheider, L. P. Kouwenhoven, and L. M. K. Vandersypen, Science {\bf 309}, 1346 (2005).
\bibitem{Nowack} K. C. Nowack, F. H. L. Koppens, Yu. V. Nazarov, and L. M. K. Vandersypen, Science {\bf 318}, 1430 (2007).
\bibitem{Tarucha} M. Pioro-Ladriere, T. Obata, Y. Tokura, Y.-S. Shin, T. Kubo, K. Yoshida, T. Taniyama, and S. Tarucha, Nature Phys. {\bf 4}, 776 (2008).
\bibitem{TaylorNP} J. M. Taylor, H.-A. Engel, W. D\"ur, A. Yacoby, C. M. Marcus, P. Zoller, and M. D. Lukin, Nature Phys. {\bf 1}, 177 (2005).
\bibitem{Hanson} R. Hanson and G. Burkard, Phys. Rev. Lett. {\bf 98}, 050502 (2007), and its auxiliary material.
\bibitem{Grover} L. K. Grover, Phys. Rev. Lett. {\bf 79}, 325 (1997).
\bibitem{Waugh} F. R. Waugh, M. J. Berry, D. J. Mar, R. M. Westervelt, K. L. Campman, and A. C. Gossard, Phys. Rev. Lett. {\bf 75}, 705 (1995).
\bibitem{Field} M. Field, C. G. Smith, M. Pepper, D. A. Ritchie, J. E. F. Frost, G. A. C. Jones, and D. G. Hasko, Phys. Rev. Lett. {\bf 70}, 1311 (1993).
\bibitem{Burkard} G. Burkard, D. Loss, and D. P. DiVincenzo, Phys. Rev. B {\bf 59}, 2070 (1999).
\bibitem{Stepanenko} D. Stepanenko, N. E. Bonesteel, D. P. DiVincenzo, G. Burkard, and D. Loss, Phys. Rev. B {\bf 68}, 115306 (2003).
\bibitem{Flindt} C. Flindt, A. S. S\o rensen, and K. Flensberg, Phys. Rev. Lett. {\bf 97}, 240501 (2006).
\bibitem{Ganga} S. Gangadharaiah, J. Sun, and O. A. Starykh, Phys. Rev. Lett. {\bf 100}, 156402 (2008).
\bibitem{Schliemann} J. Schliemann, D. Loss, and A. H. MacDonald, Phys. Rev. B {\bf 63}, 085311 (2001).
\bibitem{ae} C. Cohen-Tannoudji, J. Dupont-Roc, and G. Grynberg, {\it Atom-Photon Interactions: Basic Processes and Applications}, (Wiley, New York, 1992).
\bibitem{Bergmann} K. Bergmann, H. Theuer, and B. W. Shore, Rev. Mod. Phys. {\bf 70}, 1003 (1998).
\bibitem{Pekola1} R. Fazio, F. W. J. Hekking, and J. P. Pekola, Phys. Rev. B {\bf 68}, 054510 (2003).
\bibitem{Pekola2} M. M\"ott\"onen, J. P. Pekola, J. J. Vartiainen, V. Brosco, and F. W. J. Hekking, Phys. Rev. B {\bf 73}, 214523 (2006).
\bibitem{LZ} L. D. Landau, Phys. Z. Sowjetunion {\bf 2}, 46 (1932); C. Zener, Proc. R. Soc. A {\bf 137}, 696 (1932).
\bibitem{Ono} K. Ono, D. G. Austing, Y. Tokura, and S. Tarucha, Science {\bf 297}, 1313 (2002).
\bibitem{Miller} J. B. Miller, D. M. Zumbuhl, C. M. Marcus, Y. B. Lyanda-Geller, D. Goldhaber-Gordon, K. Campman, and A. C. Gossard, Phys. Rev. Lett. {\bf 90}, 076807 (2003).
\bibitem{Coish} W. A. Coish and D. Loss, Phys. Rev. B {\bf 72}, 125337 (2005).
\bibitem{Taylor} J. M. Taylor, J. R. Petta, A. C. Johnson, A. Yacoby, C. M. Marcus, and M. D. Lukin, Phys. Rev. B {\bf 76}, 035315 (2007).
\bibitem{Cakir} \"O. \c{C}ak\i r and T. Takagahara, Phys. Rev. B {\bf 77}, 115304 (2008).
\bibitem{Petta2} J. R. Petta, J. M. Taylor, A. C. Johnson, A. Yacoby, M. D. Lukin, C. M. Marcus, M. P. Hanson, and A. C. Gossard, Phys. Rev. Lett. {\bf 100}, 067601 (2008).
\bibitem{Kosaka} H. Kosaka, H. Shigyou, Y. Mitsumori, Y. Rikitake, H. Imamura, T. Kutsuwa, K. Arai, and K. Edamatsu, Phys. Rev. Lett. {\bf 100}, 096602 (2008).
\bibitem{Kosaka2} H. Kosaka, T. Inagaki, Y. Rikitake, H. Imamura, Y. Mitsumori, and K. Edamatsu, Nature (London) {\bf 457}, 702 (2009).
\end{thebibliography}
\end{document}